\documentclass[preprint,amsmath,amssymb,eqsecnum]{revtex4}
\usepackage{latexsym}
\usepackage{amsmath}
\usepackage{amssymb}
\usepackage{euscript}
\usepackage{epsfig}
\usepackage{rotating}
\usepackage{color}
\usepackage{hhline}
\usepackage[ps2pdf]{hyperref}
\usepackage{epsfig}
\usepackage{latexsym}

\newcommand{\sgn}{\operatorname{sgn} }

\renewcommand{\Im}{\operatorname{Im}\, }
\renewcommand{\Re}{\operatorname{Re}\, }

\begin{document} 
\title{
Effective mass of the composite fermions
and energy gaps of quantum Hall states
      }

\author{Antoine Praz} 

\affiliation{
Condensed Matter Theory Group, Paul Scherrer Institut, 
CH-5232 Villigen PSI, Switzerland
            }
\date{\today} 

%%%%%%%%%%%%%%%%%%%%%%%%%%%%%%%%%%%%%%%%%%%%%%%%%%%%%%%%%%
\begin{abstract}
The effective mass of the quasiparticles in the fermion-Chern-Simons 
description of the quantum Hall state at half-filling is computed 
for electron-electron interactions $V(r)\sim r^{x-2}$, 
for $0<x<3/2$, following the previous work
of Stern and Halperin [Phys. Rev. B {\bf 52}, 5890 (1995)].
The energy gap of quantum Hall states with filling factors $\nu=\frac{p}{2p+1}$
for $p\gg 1$ can then be obtained either from the effective mass at half-filling,
as proposed by Halperin, {\it et al.} [Phys. Rev. B {\bf 47}, 7312 (1993)],
or evaluated directly from the self-energy of the system 
in presence of the residual magnetic field;
both results are shown to agree as $p\to \infty$.
The energy gap is then given by a self-consistent equation, 
which asymptotic solution for $p\gg 1$  and short-range interactions
is $E_g(p)\sim (2p+1)^{-\frac{3-x}{2}}$,
in agreement with previous results
by Kim {\it et al.} [Phys. Rev. B {\bf 52}, 17275 (1995)].
The power law for the energy gap seems to be {\it exact}
to all orders in the perturbation expansion.
Moreover, the energy gap for systems with Coulomb interaction
is recovered in the limit $x\to 1$.
\end{abstract}

\maketitle 

\section{
Introduction
        }
\label{sec: Introduction} 

In a two-dimensional electron gas in the partially filled lowest Landau level,
a predominant role is played by the interaction between the electrons,
which sets the energy scale of the problem.
While the qualitative properties of the quantum Hall states 
are mostly unaffected by the precise form of the interaction,
the value of some quantities, as the energy gap in the incompressible Hall states, 
depends on the type of interaction.
The physical relevant interaction between the electrons is the 
electrostatic Coulomb interaction.
However, the study of systems having more 
general interactions of long  or short ranges between the fermions is of great interest. 
In this work, we will consider the quantum Hall system at or near half-filling
with different types of electron-electron interactions,
using the fermion-Chern-Simons (FCS) approach.~\cite{Halperin03}

The Chern-Simons composite fermions are constructed attaching
an even numbers $\phi$ of flux quanta to the electrons;~\cite{Jain89}
this transformation can be realized introducing an appropriated 
Chern-Simons gauge field.~\cite{Lopez91}
As a consequence, the external magnetic field is partially screened,
but a long ranged, gauge-mediated interaction between the new fermions 
has been introduced.

At the mean-field level, the system can be described in terms 
of free composite fermions in a constant effective magnetic field
$\Delta B = B\big(1 - \phi \nu\big)$,
where $\nu=2\pi n \hbar c/(e B)$ is the filling factor and $n$ the electron density.
Choosing $\phi=2$ at half-filling ($\nu=1/2$) gives $\Delta B = 0$;
the ground state of the system is then the Fermi sea of composite fermions, 
with Fermi momentum $k_F = \sqrt{4\pi n}$.
Along the main Jain series, which is determined by the filling factors 
$\nu = \frac{p}{2p+1}$ for integer $p$, the fractional quantum Hall states 
correspond to the integer quantum Hall states for composite fermions
in the residual magnetic field $\Delta B = B/(2p+1)$, 
with $p$ Landau levels filled.

More generally, the mean-field picture predicts
the occurrence of fractional quantum Hall states
at filling factors that corresponds to integer filling factors
for the composite Fermions.
These predictions are believed to be correct,
providing that the interaction belongs to 
a well-defined class of ``hard-core'' repulsive interactions.~\cite{Wojs}
In this case, the qualitative properties of the quantum Hall states
do not depend on the precise form of the interactions.

We are interested in the energy gap of the incompressible quantum Hall state at
$\nu = \frac{p}{2p+1}$, which corresponds to the energy required
to create a quasiparticle and a quasihole infinitely far apart.
In the mean-field approximation, the energy gap 
is given by the separation of the Landau energy levels in the residual magnetic field 
$\Delta B=\frac{B}{2p+1}$. In particular, 
the mean-field energy gap at $\nu = \frac{p}{2p+1}$
falls off as $E_{mf}(p)\sim 1/(2p+1)$ in the limit of large $p$;
however, the energy scale of the mean-field gap is set
by the cyclotron energy $\hbar eB/(cm)$, where $m$ is the band mass of the electrons,
and not by the interaction energy as expected.

In order to include the effects of gauge fluctuations, %
Halperin, Lee and Read (HLR)~\cite{HLR} conjectured that the energy gap 
of the quantum Hall states belonging to the main Jain series
is given by the cyclotron energy of the composite fermions in the
residual magnetic field $\Delta B$, with the bare mass of the quasiparticles
replaced by their effective mass $m^*(\omega)$ at the gap energy $E_g(\nu)$,
\begin{equation}
E_g(\nu) = \frac{\hbar e |\Delta B| }{c\, m^*\big(E_g(\nu)\big)}.
\label{eq: HLR conjecture}
\end{equation}
This equation for the energy gap will be referred to as the HLR-gap equation.

Halperin {\it et al}. have computed in their seminal paper~\cite{HLR}
the effective mass of the composite fermions for various interactions,
including the gauge fluctuations up to the one-loop level.
They concluded that the effective mass 
of the composite fermions diverges at the Fermi level
for short range and Coulomb interactions,
while remaining finite for long range interactions.
In the case of the Coulomb interaction, the effective mass of the
composite fermions is given by~\cite{HLR,Stern95}
\begin{equation}
m^*_{\mathrm{Cb}}=
\lim_{\omega\to 0}\frac{2}{\pi}\frac{\varepsilon k_F}{e^2}|\ln\omega| + \cdots,
\label{eq: effective mass Coulomb}
\end{equation}
where $\varepsilon$ is the dielectric constant, and the dots represent subleading terms.
Stern and Halperin~\cite{Stern95} argued that this 
result is exact to all orders in perturbation theory,
up to finite contributions from short wavelengths.
For short range interactions of the form $V(r)\sim 1/r^{2-x}$ with $0<x<1$, 
the divergence of the effective mass obeys the power law~\cite{HLR,Kim95}
\begin{equation}
m^*_{\mathrm{sr}}\sim \lim_{\omega\to 0}\omega^{\frac{x-1}{3-x}}.
\end{equation} 
Moreover, in the one-loop approximation, 
the imaginary part of the effective mass $m^*_{\mathrm{sr}}$ 
diverges with the same rate as its real part does.
These results suggest that, at half-filling, 
the Fermi velocity $v_F = k_F/m^*$ goes to zero, 
and that the lifetime of the quasiparticles is vanishingly short.~\cite{Kim95}
In particular, the Fermi liquid description
of the compressible Hall states at half-filling 
with short range interaction is expected to break down.
This picture is consistent with the renormalization group analysis
proposed by Nayak and Wilczek.~\cite{Nayak94}
Indeed, using an approach inspired by the $\epsilon$ expansion,
Nayak and Wilczek 
showed that the Fermi liquid description of the fermion-Chern-Simons theory 
is a fixed point of the renormalization group flow as long as the interaction is long ranged.
For short range interactions, the theory flows aways from the Fermi liquid 
fixed point toward strong coupling,
eventually leading to a breakdown of the Fermi liquid description.
The theory with Coulomb interaction is marginal; 
it flows toward the weak-interacting fixed point, 
but the two-point correlation function for the composite fermions
exhibits at low energy a non-Fermi liquid behavior $G(\omega)\sim 1/(\omega\ln\omega)$.

The question of the physical meaning of the divergent effective mass 
has to be addressed in terms of gauge-invariant quantities.
For instance, the singular self-energy leading to a divergence of the effective mass
is a gauge-dependent quantity.
In the gauge-invariant density-density and current-current correlation functions,
important cancellations between the self-energy corrections and the vertex corrections
due to Ward identities lead to a Fermi liquid behavior for all ratios of $\omega$ 
and $k_F |\boldsymbol{k}|/m$.~\cite{Kim94}
However, other gauge-invariant quantities such as the specific heat~\cite{Kim96}
are sensitive to the singularity in the effective mass.
In this context, Kim and Lee~\cite{Kim96} showed that for the Coulomb interaction, 
the singular part of the specific heat of the composite fermion system 
predicted by the quasiparticle approximation coincides with
the singular part of the specific heat obtained from the free energy of the gauge field.
For short range interactions, both results for the specific heat disagree,
and they concluded that the Landau quasiparticle approximation is valid only
for the Coulomb interaction.

Aware of the limitations of the Fermi liquid approach for short range interactions,
we might still try to use it to compute some properties of the incompressible Hall states near 
half-filling. 
In particular, we may hope that the quasiparticle approximation 
will make valid qualitative predictions for the behavior of the energy 
gap along the main Jain series as function of $p$.

For the Coulomb interaction, the HLR-gap equation~(\ref{eq: HLR conjecture}) has been confirmed 
by one-loop calculations in the vicinity of half-filling 
by Stern and Halperin.~\cite{Stern95}
According to formula~(\ref{eq: HLR conjecture}), 
the energy gap of the quantum Hall states at the filling fraction $\nu=\frac{p}{2p+1}$
in the limit of large $p$ behaves as
\begin{equation}
E_g(\nu) \simeq
\frac{\pi}{2}\frac{e^2k_F}{\varepsilon}\frac{1}{(2p+1)\big[\ln (2p+1)+C'\big]},
\label{eq: energy gap Coulomb}
\end{equation}
where the constant $C'$ depends on short wavelength contributions,
and cannot be evaluated within the fermion-Chern-Simons theory.
However, the coefficient of the logarithmic term is believed to be exact.

Numerical studies of the energy gap in the incompressible quantum Hall states
with $\nu=1/3$, $2/5$, $3/7$ and $4/9$ (corresponding to $p=1,2,3,$ and $4$)
have been performed by Morf {\it et al.}~\cite{Morf02}
for the Coulomb interaction.
They have computed the energy gap of systems of up to 18 electrons on 
a sphere, and extrapolated the result to the limit of infinite systems.
Using $C'$ as a fitting parameter, they found a good agreement between 
formula~(\ref{eq: energy gap Coulomb}) and the numerical results.
Moreover, the fit with Eq.~(\ref{eq: energy gap Coulomb}) is better
than it would be without the logarithmic correction.

These results were a motivation to produce a counterpart to Eq.~(\ref{eq: energy gap Coulomb})
for long and short range interactions, in view of comparison with numerical studies.
For this purpose, we have reproduced the study of Stern and Halperin~\cite{Stern95} 
for general interactions, in spite of the strong evidence for
the breakdown of the Fermi liquid approach when the interaction is 
short ranged. 
Following Ref.~\onlinecite{Stern95}, we first compute the
effective mass at half-filling
in the presence of an electron-electron interaction determined by the potential
\begin{equation}
V(r)\sim \frac{1}{r^{2-x}},
\label{eq: potential r-space}
\end{equation}
where $x<1$ corresponds to short range interaction, and $x=1$ is the Coulomb interaction,
while for $x>1$, the interaction is long ranged.%
~\cite{footnote: pseudo-potentials}
For $0<x<1$, we find the effective mass
\begin{equation}
m^* \simeq
\frac{\phi^2 F(x)}{\hat{V}(k_F)}
\lim_{\omega\to 0} 
\left(\frac{4\pi \phi \omega}{k_F^2\hat{V}(k_F)}\right)^{(x-1)/(3-x)},
\label{eq: effective mass short range}
\end{equation}
in agreement with the effective mass previously obtained in the literature.~\cite{HLR,Kim95}
Here, $\hat{V}(k)\sim k^{-x}$ 
is the Fourier transform of the potential~(\ref{eq: potential r-space})
and $k_F=\sqrt{4\pi n}$ the Fermi momentum. 
$F(x)$ is a complex function, with a real part that diverges at $x= 1$.
The effective mass $m^*$ diverges following a power law for $x<1$;
in the limit $x\to 1$, the effective mass~(\ref{eq: effective mass Coulomb})
obtained for the Coulomb interaction is recovered.
Using an argument based on Ward identities to account for renormalization
of the vertex function, we argue that Eq.~(\ref{eq: effective mass short range})
is exact; however, unlike in the case of Coulomb interaction,
the function $F(x)$ may get contributions from all orders in the loop expansion.
In other words, the contributions from higher orders 
do not modify the strength of the divergence. 

In a second step, we have computed the energy gap at the incompressible quantum 
Hall states at $\nu = \frac{p}{2p+1}$ and in the limit of large $p$, 
still following Ref.~\onlinecite{Stern95}.
The self-consistent equation for the energy gap obtained in presence of
the effective magnetic field $\Delta B$ corresponds
to the HLR-formula~(\ref{eq: HLR conjecture}), up to subleading corrections.
The asymptotic behavior of the energy gap for large $p$ is then
\begin{equation}
E_g(p) \simeq
Y(x)
\frac{k_F^2\hat{V}(k_F)}{4\pi \phi}
\left(\frac{2\pi}{2p+1} \right)^{(3-x)/2}.
\label{eq: energy gap short range}
\end{equation}
In the one-loop approximation, $Y(x)$ is a complex function of $x$,
and the solution of the gap equation~(\ref{eq: HLR conjecture})
has a nonvanishing, unphysical imaginary part, which comes 
probably from the fact that higher order terms in the loop expansion have been neglected.

In the case of Coulomb interaction, the imaginary part obtained solving the
one-loop gap equation~(\ref{eq: HLR conjecture}) is subleading, and vanishes
faster than the real part as $p$ tends to infinity; 
hence, for the Coulomb interaction, the energy gap obtained 
at the one-loop order is asymptotically exact in the limit $p\to \infty$.
For short range interactions, both the real and imaginary parts of the 
solution of Eq.~(\ref{eq: HLR conjecture}) behave the same way in the limit $p\to \infty$. 
The coefficient $Y(x)$ in Eq.~(\ref{eq: energy gap short range}) remains,
therefore, undetermined at the one-loop order for any $p$, 
and only the overall power-law behavior of the energy gap 
as function of $p$ is captured by the one-loop calculations.

We interpret the appearance of imaginary parts both
in the effective mass and in the energy gap as a signature
of the breakdown of the composite Fermion approach.
The loop expansion relies on the smallness of the 
gauge fluctuations around its mean-field value.
For long range interactions, long wavelength gauge fluctuations
are suppressed, and the FCS approach is justified;
the effective mass and energy gap are finite and real.
However, since both long  and short wavelength contributions 
are equally important, the FCS approach is not appropriated
to give an approximation for the effective mass or the energy gap
in this case.

For short range interactions, the gauge field may fluctuate
strongly, eventually leading to the breakdown of the FCS picture;
any prediction has, therefore, to include all orders in the loop expansion.
In doing so, we could compute the power-law behavior of the 
most divergent part of the effective mass, and the $p$ dependency
of the energy gap; the contributions coming from short wavelengths
are subleading, and may be neglected in this context. 
The Coulomb interaction is marginal:
the long wavelength gauge fluctuations are sufficiently suppressed
by the long range interaction to allow loop expansion,
but remain strong enough to produce a divergence in the effective mass,
which can then be obtained exactly by a one-loop calculation
including only long wavelength fluctuations.

This paper is constructed as follows. A brief description
of the fermion Chern-Simons theory is given in Sec.~\ref{sec: The Fermion-Chern-Simons Theory}.
The effective mass is then computed at half-filling from the one-loop approximation
of the self-energy, whereby the role of higher-order contributions in the loop expansion
is also discussed. In Sec.~\ref{sec: Evaluation of the Energy Gap},
the gap equation and its solution are discussed.

%%%%%%%%%%%%%%%%%%%%%%%%%%%%%%%%%%%%%%%%%%%%%%%%%%%%%%%%%%
\section{Fermion-Chern-Simons Theory}
\label{sec: The Fermion-Chern-Simons Theory}

In this section, we briefly review the construction of the
fermion-Chern-Simons theory. 
We start from the zero-temperature partition function for 
a two-dimensional electron gas in a strong magnetic field $B$
perpendicular to the plane, which is (in units where $\hbar=c=1$)
\begin{equation}
Z = \int \mathcal{D}[\psi^+_e,\psi_e] e^{iS},
\end{equation}
where $\psi_e^+$ and $\psi_e$ are the anticommuting electron fields.
The action in the real-time formalism is given by
\begin{subequations}
\label{eq: definition of action}
\begin{equation}
S = \int dt \int d^2r\, 
\big(\mathcal{L}_0+\mathcal{V}\big),
\end{equation}
with the Lagrangian density
\begin{eqnarray}
\mathcal{L}_0 & = &
\psi_e^+
\left[
i\partial_t-\frac{1}{2m}\big(-i\boldsymbol{\nabla}+e\boldsymbol{A}\big)^2+\epsilon_F
\right]
\psi_e,
\\
\mathcal{V} & = & 
\frac{1}{2}\int d^2r'\,
\rho_e(t,\boldsymbol{r})
V(\boldsymbol{r}-\boldsymbol{r}')
\rho_e(t,\boldsymbol{r}'),
\end{eqnarray}
\end{subequations}
where $\epsilon_F$ is the chemical potential,
$m$ is the band mass of the electrons,
$\boldsymbol{A}$ is the vector potential corresponding to  
the magnetic field $B=\nabla\wedge \boldsymbol{A}$ and
$\rho_e(t,\boldsymbol{r})=\psi_e^+(t,\boldsymbol{r})\psi_e(t,\boldsymbol{r})$
is the electron density.
The precise form of the static electron-electron interaction potential $V$ 
will be discussed later.

We introduce the composite fermions $\psi$, which are related to
the electron fields by a singular gauge transformation,
\begin{equation}
\label{eq: definition composite fermions}
\psi^+ =
\psi_e^+
\exp\left[
-i\phi\int d^2r' \arg(\boldsymbol{r}-\boldsymbol{r}')\rho_e(t,\boldsymbol{r}') 
\right],
\end{equation}
where $\arg(\boldsymbol{r})$ is the angle between the vector $\boldsymbol{r}$ and
the $x$ axis. 
The fields $\psi$ and $\psi^+$ satisfy the usual anticommutation
relations, providing $\phi$ is an even integer; we will specialize to the case $\phi=2$.
The composite fermion density remains the same as the electron density,
\begin{equation}
\rho(t,\boldsymbol{x}) =
\psi^+(t,\boldsymbol{x})\psi(t,\boldsymbol{x})
= \rho_e(t,\boldsymbol{x}).
\end{equation}
This transformation is associated with the gauge potential,
\begin{subequations}
\begin{eqnarray}
\boldsymbol{a} & = & 
\phi \int d^2r' \boldsymbol{g}(\boldsymbol{r}-\boldsymbol{r}')\rho(t,\boldsymbol{r}'),
\\
\boldsymbol{g}(\boldsymbol{r}) & = & 
( \hat{z}\times \boldsymbol{r} )/|\boldsymbol{r}|^2.
\end{eqnarray}
\end{subequations}
Observe that the vector potential $\boldsymbol{a}$ satisfies to
the constraint
\begin{eqnarray}
\boldsymbol{\nabla} \times \boldsymbol{a} & = &
2\phi \pi \rho(t,\boldsymbol{r}),
\label{eq: constraint on CS gauge}
\end{eqnarray} 
where $\boldsymbol{\nabla}\times\boldsymbol{a}=\epsilon^{ij}\partial_i a_j$.
We choose to work in the Coulomb gauge, 
with $\boldsymbol{\nabla}\cdot\boldsymbol{a}=0$. The longitudinal 
part of the gauge vector vanishes, and we denote $a_\perp$ its transversal part,
satisfying $a_i = \epsilon^{ij} k_j a_\perp/|\boldsymbol{k}|$ in momentum space.
The constraint~(\ref{eq: constraint on CS gauge}) can be ensured by the
Lagrange multiplier field $a_0$, adding the term
\begin{equation}
 \int dt \int d^2r\,   a_0 
\left(
\frac{1}{2\phi \pi }\boldsymbol{\nabla} \times \boldsymbol{a} 
-\rho(t,\boldsymbol{r}) 
\right)
\end{equation}
to the action. The partition function can be rewritten in terms of the
composite fermion and gauge fields as
\begin{equation}
\label{eq: FCS partition function}
Z =
\int \mathcal{D}[a_0,a_\perp]\int \mathcal{D}[\psi^+,\psi] e^{iS_{\mathrm{FCS}}}
\end{equation}
where the fermion-Chern-Simons action is obtained from the
the Lagrangian density, 
\begin{subequations}
\begin{equation}
\label{eq: FCS Lagrangian}
\mathcal{L}_{\mathrm{FCS}}=\mathcal{L}_{\mathrm{CF}}+\mathcal{L}_{\mathrm{CS}}+\mathcal{V}.
\end{equation}
The Lagrangian for the composite fermions is
\begin{equation}
\mathcal{L}_{\mathrm{CF}} = 
\psi^+ \left[
i\partial_t - a_0
-\frac{1}{2m}\big(-i\boldsymbol{\nabla}-\boldsymbol{a}+e\boldsymbol{A}\big)^2
+\epsilon_F
       \right]  \psi
\end{equation}
and the Chern-Simons Lagrangian for the gauge field in the Coulomb gauge
\begin{equation}
\mathcal{L}_{\mathrm{CS}}
=\frac{1}{2\phi\pi}\epsilon^{ij}  a_0 \partial_i a_j.
\end{equation}
\end{subequations}
In the mean-field approximation, the density of electrons is assumed
to be constant, such that, in virtue of Eq.~(\ref{eq: constraint on CS gauge}),
the composite fermions are subject to the reduced magnetic field
\begin{equation}
\begin{split}
\Delta B & = B - 2\pi\phi n/e
\\            & = B\big( 1-\phi \nu \big),
\end{split}
\end{equation}
where $n=\langle \rho\rangle$ is the mean density of particles
and $\nu = 2\pi n /(e B)$ the filling factor.
In particular, at half-filling ($\nu=1/2$) 
the residual magnetic field vanishes providing $\phi = 2$.
Along the main Jain series determined by $\nu=p/(2 p+1)$,
the effective magnetic field becomes
\begin{equation}
\Delta B^{(p)} = \frac{B}{2 p+1}.
\end{equation}
According to the mean-field predictions, the quantum Hall 
system at half-filling is in a compressible state,
which may be described in terms of a Fermi liquid of composite fermions;
furthermore, the incompressible quantum Hall states of the main Jain series
are described in terms of the integer quantum Hall effect for composite fermions.

In order to go beyond the mean-field approximation, one has to take into
account fluctuations of the gauge field.
For this purpose, we expand the Lagrangian~(\ref{eq: FCS Lagrangian})
up to the second order in fluctuations of the Chern-Simons gauge field $(a_0,a_\perp)$ 
around its mean-field value. Doing so, and using 
the constraint~(\ref{eq: constraint on CS gauge})
in order to express the interaction term $\mathcal{V}$ of Eq.~(\ref{eq: FCS Lagrangian}) 
solely in terms of the gauge field, we obtain the effective action
\begin{equation}
\mathcal{S}_{\mathrm{eff}} =
\int dt \int d^2r\, \mathcal{L}_{\mathrm{CF}} 
+ \mathcal{S}_{\mathrm{gauge}},
\end{equation}
where in momentum space,
\begin{equation}
\mathcal{S}_{\mathrm{gauge}} =
\frac{1}{2}\int \frac{d^3k}{(2\pi)^3}
a_\lambda(-\boldsymbol{k},-\omega)
\big(D^{-1}_0\big)_{\lambda\mu}
a_\nu (\boldsymbol{k},\omega).
\end{equation}
The Greek indices run over the longitudinal and perpendicular directions, and
the inverse bare gauge propagator is
\begin{equation}
D^{-1}_0   = 
\begin{pmatrix}
0  & \frac{ik}{2\phi\pi}
\\
-\frac{ik}{2\phi\pi} &  -\frac{k^2 \hat{V}(k) }{2\pi\phi^2}
\end{pmatrix}.
\label{eq: bare gauge propagator}
\end{equation}
We will consider electron-electron interactions of the from
\begin{equation}
\hat{V}(k) =
\frac{\hat{V}(k_F)}{(k/k_F)^x} 
\end{equation}
in momentum space, with $3/2 > x > 0$.~\cite{footnote: fourier trafo V} 
For the Coulomb interaction, $x=1$ and $k_F\hat{V}(k_F)=2\pi e^2/\varepsilon$
with the dielectric constant $\varepsilon$;
$x<1$ corresponds to short range interactions.
For $x>1$, the interaction is long range, i.e., falls off 
slower than $1/r$ in position space. 
We assume the occurrence of the quantum Hall states for all
filling factors in the Jain series; this is reasonable since
the corresponding Haldane pseudopotentials
exhibit hard-core-like properties,~\cite{Wojs}
as shown in Appendix~\ref{app: pseudo-potentials}. 

We are interested in the effective mass for the composite fermions,
which is given in terms of the self-energy by
\begin{equation}
m^* =
m \left.
\frac{
1-\partial_\omega \Sigma}
     {1+\partial_{ \epsilon_{\boldsymbol{k}}} \Sigma}
\right|_{\omega=0, k=k_F}
\label{eq: definition of eff. mass} 
\end{equation}
where $\epsilon_{\boldsymbol{k}}=\boldsymbol{k}^2/(2m)-\epsilon_F$ 
is the bare dispersion relation.

%%%%%%%%%%%%%%%%%%%%%%%%%%%%%%%%%%%%%%%%%%%%%%%%%%%%%%%%%%
\section{Effective mass of the composite fermions}

We turn to the computation of the self-energy to the one-loop order
(first order in the gauge propagator $D$). 
In the following calculations, the gauge propagator $D$ in the random phase
approximation~\cite{HLR,Stern95} will be used.
In the Coulomb gauge, using the $(0,\perp)$-notation introduced above,
\begin{equation}
D^{-1}(\omega,\boldsymbol{k}) =
\begin{pmatrix}
\frac{m}{2\pi} & \frac{i|\boldsymbol{k}|}{2\pi\phi}
\\
-\frac{i|\boldsymbol{k}|}{2\pi\phi}\, &
i\frac{2n}{k_F}\frac{\omega}{k}
-\boldsymbol{k}^2\big(\frac{1}{12\pi m}
+\frac{k_F^x \hat{V}(k_F)}{(2\pi \phi)^2 |\boldsymbol{k}|^x}\big)
\end{pmatrix}.
\label{eq: RPA gauge propagator}
\end{equation}
The one-loop computations of the self-energy involve
only the diagonal part of the gauge propagator. 
In the limit of long wavelengths $\boldsymbol{k}\to 0$
and small frequencies $\omega\to 0$, 
\begin{equation}
D_{\perp\perp}(\omega,\boldsymbol{k}) \simeq
\frac{\phi \pi}{\omega_{\mathrm{int}}} \frac{ |\boldsymbol{k}|}{k_F}
\left[
i \frac{\omega}{\omega_{\mathrm{int}}} 
- \left(\frac{|\boldsymbol{k}|}{k_F}\right)^{3-x}
\right]^{-1}
\end{equation}
where for $x>0$,~\cite{footnote: short range interaction}
\begin{equation}
\label{eq: definition of interaction energy}
\omega_{\mathrm{int}} =
k_F^2\hat{V}(k_F)/\big(4\pi \phi\big)
\end{equation}
is the energy scale defined by the electron-electron interaction; 
for $x=1$, $\omega_{\mathrm{int}}= e^2 k_F /(2\phi\varepsilon)$.
Further, the longitudinal gauge propagator is $D_{00}\simeq 2\pi/m$.

%%%%%%%%%%%%%%%%%%%%%%%%%%%%%%%%%
\subsection{Self-energy}

To the first order in $D$, the most important contribution
to the effective mass comes from the transversal part 
of the self-energy 
\begin{equation}
\begin{split}
\Sigma_\perp(\omega,\boldsymbol{k}) = 
\frac{i}{m^2} \int & \frac{d\Omega}{2\pi} \int \frac{d^2 k'}{(2\pi)^2} 
\frac{(\boldsymbol{k}\times\boldsymbol{k}')^2}{(\boldsymbol{k}-\boldsymbol{k}')^2}
\\
& \times
D_{\perp\perp}(\Omega,\boldsymbol{k}-\boldsymbol{k}')
G_0(\omega-\Omega,\boldsymbol{k}'),
\end{split}
\label{eq: transversal self-energy}
\end{equation}
where 
\begin{equation}
G_0(\omega,\boldsymbol{k}) =
\frac{1}{\omega -\epsilon_{\boldsymbol{k}}+ i0_+ \sgn(\omega)}
\label{eq: bare fermion propagator}
\end{equation}
is the free propagator for the composite fermions at half-filling,
with $\epsilon_{\boldsymbol{k}}=\boldsymbol{k}^2/(2m)-\epsilon_F$.
The self-energy~(\ref{eq: transversal self-energy})
is computed in the same way as in the case of electron-phonon coupling.~\cite{Lifshitz}
We compute
$\Sigma_\perp(\omega,\boldsymbol{k})-\Sigma_\perp(0,\boldsymbol{k})$
and introduce the variable $q=|\boldsymbol{k}-\boldsymbol{k}'|$,
changing the coordinates according to
\begin{equation}
\int d^2 k' f(\boldsymbol{k}')
=
\int_0^\infty dk'
\int_{|k-k'|}^{k'+k} 
\frac{dq\, k' q}{|\boldsymbol{k}\times\boldsymbol{k}'|}
f(k',q).
\end{equation}
The dominant contribution to the transversal self-energy comes from 
the region of the momentum space where $k'\simeq k_F$,
due to the pole in the fermion propagator~(\ref{eq: bare fermion propagator});
we thus neglect the $k'$-dependency of the integrand, except in 
$G_0(\Omega-\omega,\boldsymbol{k}')$.
Furthermore, the $k$ dependency of the self-energy 
leads to subleading corrections, and we may assume $k\simeq k_F$.
The integration over $k'$ is, finally, performed closing the contour
in the upper complex half-plane. As a consequence, the integration
over the frequency $\Omega$ is restricted to the range $[0,\omega]$.
Integrating $D_{\perp\perp}(\Omega,q)$ over $\Omega$ leads to
\begin{equation}
\Sigma_\perp(\omega,k_F) -\Sigma_\perp(0,k_F) 
\simeq
-\frac{\phi k_F^2 }{2\pi m }\frac{I\big(2u_{\mathrm{max}}\big)}{u_{\mathrm{max}}^2},
\label{eq: one-loop transversal self-energy, intermed. step}
\end{equation}
where we have define the integral
\begin{equation}
I(v) =
\int_0^{v}\!\! du\, u
\left\{ i\sgn(\omega)\ln \sqrt{1+u^{2x-6}}
+
\arctan u^{x-3}
\right\}.
\label{eq: integral I(v)}
\end{equation}
The dimensionless variable $u$ is given by
\begin{equation}
u = \frac{q}{k_F} \left(\frac{\omega_{\mathrm{int}}}{|\omega|}
\right)^{1/(3-x)},
\label{eq: definition of u}
\end{equation}
with 
$u_{\mathrm{max}} = \big(\omega_{\mathrm{int}}/|\omega|\big)^{1/(3-x)}$. 
For $\omega\to 0$, the integral~(\ref{eq: integral I(v)})
has to be evaluated for $v\gg 1$, as done in Appendix~\ref{app sec: The integral I(v)}.
Inserting the approximation~(\ref{app eq: I(v) for large v}) for $I(v)$
in the expression~(\ref{eq: one-loop transversal self-energy, intermed. step}) 
for the self-energy leads to
\begin{equation}
\begin{split}
& \Sigma_\perp(\omega,k_F) -\Sigma_\perp(0,k_F) 
\\
& \simeq 
|\omega|\frac{m_{\mathrm{int}}}{m }
\left\{
F_0(x)\big(|\omega|/\omega_{\mathrm{int}}\big)^{(x-1)/(3-x)}
+\frac{2^{x}}{x-1}
+\cdots
\right\}
\end{split}
\label{eq: one-loop self-energy, all x}
\end{equation}
where we have defined the interacting mass scale
\begin{equation}
m_{\mathrm{int}} = \frac{\phi k_F^2}{4\pi \omega_{int} }.
\label{eq: definition of m_int}
\end{equation} 
For the Coulomb interaction, $m_{\mathrm{int}} = \frac{\phi^2 \varepsilon k_F}{2\pi e^2}$.
The function $F_0(x)$ is obtained from the integral~(\ref{eq: integral I(v)})
and is given by 
\begin{eqnarray}
\Re F_0(x) & = &
\frac{\pi(3-x)}{4}
\nonumber
\\
&-&
\sum_{n\ge 0}\frac{4(3-x) (-1)^n}{(2n+1)\big[4-(2n+1)^2(3-x)^2\big]}
\nonumber
\\
\Im F_0(x) & =&
\sgn(\omega)(3-x)\int_0^\infty \frac{u\, du}{1+u^{6-2x}}.
\label{eq: definition of F0}
\end{eqnarray}
Notice that the real part of $F_0(x)$ diverges as $x\to 1$, while its imaginary part 
remains finite.
For short range interactions ($x<1$),
both the real and imaginary parts of the self-energy 
are proportional to $|\omega|^{2/(3-x)}$.
In particular, the Landau criterion for the existence of the
quasiparticle is not fulfilled, since
$|\Im\Sigma(\omega,\boldsymbol{k})|> |\,\omega|$ for $\omega\to 0$.
The prediction of the Fermi liquid theory 
should, therefore, be considered with care if the interaction between
the electrons is short ranged, as already discussed in Ref.~\onlinecite{Kim96}.
For $x< 1$, the leading term of the 
self-energy~(\ref{eq: one-loop self-energy, all x}) comes 
from the integration region where  
$u \lesssim 1$ in Eq.~(\ref{eq: integral I(v)});
this corresponds to momenta $q$ smaller than $q_0\sim |\omega|^{\frac{1}{3-x}}$.
In other words, for systems with short range interactions,
the divergence of the effective mass comes from small frequencies and small momenta.

In the limit $x\rightarrow 1$, the function $F_0(x)$ can be replaced
by its limiting value
\begin{equation}
F_0(x\to 1) =
-\frac{2}{x-1}
+i\frac{\pi}{2}+2,
\end{equation}
and Eq.~(\ref{eq: one-loop self-energy, all x}) becomes
\begin{equation}
\begin{split}
&\Sigma_\perp(\omega,k_F) -\Sigma_\perp(0,k_F) \simeq 
\\
&-|\omega|\frac{m_{\mathrm{int}}}{m}
\left\{
-\frac{2}{x-1}
\left(\big(|\omega|/\omega_{\mathrm{int}}\big)^{(x-1)/2}-1 \right)
+i\sgn(\omega)\frac{\pi}{2}
\right\}.
\label{eq: one-loop self-energy, x->1}
\end{split}
\end{equation}
Expanding in powers of $x-1$ leads to the 
logarithmic behavior characteristic for the 
Coulomb interaction,~\cite{HLR,Stern95}
\begin{equation}
\begin{split}
\Sigma_\perp(\omega,k_F) & -\Sigma_\perp(0,k_F) 
 \simeq \\
& -\frac{\phi^2}{2\pi}\frac{\varepsilon k_F }{e^2 m }
|\omega|
\left(
\ln\frac{k_F e^2 }{2\phi\varepsilon |\omega|}+i\sgn(\omega)\frac{\pi}{2}
\right).
\end{split}
\end{equation}
The logarithm is obtained combining the
leading term of Eq.~(\ref{eq: one-loop self-energy, x->1}),
which comes from integration over small momenta $q$,
with the subleading term $\sim 2\omega/(x-1)$, 
coming from integration over large $q$ in Eq.~(\ref{eq: integral I(v)}).
Hence, as pointed out in Ref.~\onlinecite{Stern95},
the logarithmic part of the self-energy comes
from small frequencies, but from a large range of momenta.
The imaginary part of the self-energy is much smaller than its real part
in the limit $\omega\to 0$,
justifying the quasiparticle approach for systems with Coulomb interaction.

Finally, for long range interactions ($x>1$), the self-energy is proportional to 
$\omega$, and its imaginary part is of higher-order in $\omega$. 
The Fermi liquid picture is thus well justified in this case,
as pointed out by Nayak and Wilczek.~\cite{Nayak94}

%%%%%%%%%%%%%%%%%%%%%%%%%%%%%%%%%%%%%%%%%%%%%%%%%%%%%%%%%%
\subsection{Effective mass}
\label{ssec: Effective mass}

We may now compute the effective mass in the one-loop approximation from the
self-energy obtained in Eq.~(\ref{eq: one-loop self-energy, all x}).
We will consider only the most divergent part of the effective mass,
neglecting finite contributions from the momentum dependency of the 
self-energy or from the longitudinal part of the self-energy.
Using formula~(\ref{eq: definition of eff. mass}), we
get for short range interactions in the one-loop approximation
\begin{equation}
\frac{m^*}{m_{\mathrm{int}}} \simeq
F_0(x)
\lim_{\omega\to 0} \left(\frac{\omega}{\omega_{\mathrm{int}}}
\right)^{(x-1)/(3-x)}
+\frac{2^{x}}{x-1}
+\cdots,
\label{eq: effective mass short range 2}
\end{equation}
where $m_{\mathrm{int}}$ was defined in Eq.~(\ref{eq: definition of m_int}).
The effective mass thus diverges following a power-law on the Fermi surface,
according to the prediction of Ref.~\onlinecite{HLR}.
Moreover, the imaginary part of the effective mass diverges,
leading to instabilities in the Fermi liquid at half-filling with short range
interactions.
Again, we recover the previous results
by Halperin and Stern~\cite{Stern95} in the limit $x\to 1$, 
\begin{equation}
m^* =
\frac{\phi^2}{2\pi}\frac{\varepsilon k_F }{e^2 }
\left(
\ln\frac{k_F e^2 }{2\phi\varepsilon \omega}+i\frac{\pi}{2}
\right)+\mathrm{const}.
\label{eq: one-loop result for effective mass x=1}
\end{equation}
For long range interactions ($x>1$), we expect the effective mass to be 
real and finite; in this case, higher-order terms and 
short wavelengths contributions have to be taken into account in order to 
determine the value of the effective mass.

%%%%%%%%%%%%%%%%%%%%%%%%%%%%%%%%%%%%%%%%%%%%%%%%%%%%%%%%%%
\subsection{Beyond the one-loop approximation}
\label{ssec: Beyond the one-loop approximation}

Since there is no small expansion parameter, the result of 
the field theory can be trusted only if all orders in the perturbation theory
can be controlled. 
In the case of Coulomb interaction,
it is believed that the divergent part of the 
effective mass~(\ref{eq: one-loop result for effective mass x=1})
is \textit{exact} to all orders in the perturbation theory.~\cite{Stern95}
We will discuss here the effect of higher-order contributions
on the divergence of the effective mass for general interactions,
and eventually argue that the power-law divergence in 
Eq.~(\ref{eq: effective mass short range 2}) is the same to all orders for $x< 1$, 
whereby the one-loop function $F_0(x)$ defined in Eq.~(\ref{eq: definition of F0}) 
has to be replaced by its renormalized counterpart $F(x)$.
We use an argument relying on Ward identities 
similar to the argument used by Halperin and Stern~\cite{Stern95} 
for systems with a Coulomb interaction.

%%%%%%%%%%%%%%%%%%%%%%%%%%%%%%%%%
\subsubsection{Self-consistent equation for the self-energy}

A first improvement to the one-loop calculations can be obtained
considering a self-consistent equation for the self-energy,
replacing in Eq.~(\ref{eq: transversal self-energy}) 
the bare fermion propagator~(\ref{eq: bare fermion propagator}) by
\begin{equation}
G(\omega,\boldsymbol{k}) =
\frac{1}{G_0^{-1}(\omega,\boldsymbol{k})
-\Sigma(\omega,\boldsymbol{k})}.
\label{eq: RPA propagator}
\end{equation}
The self-energy obtained is then the sum over rainbow diagrams.
Repeating the calculations that led to the one-loop 
result~(\ref{eq: one-loop self-energy, all x}) shows that the leading
term of the self-energy, and, consequently, the divergent part of the 
effective mass~(\ref{eq: effective mass short range 2}),
is not affected by this procedure; the function $F_0(x)$ defined in 
Eq.~(\ref{eq: definition of F0}) is not modified either.

%%%%%%%%%%%%%%%%%%%%%%%%%%%%%%%%%
\subsubsection{Short wavelength contributions}

Following Ref.~\onlinecite{Stern95}, we now show that the short wavelength contributions
to the self-energy do not modify the power-law behavior of the most divergent part 
of the effective mass.
For this purpose, we decompose the momentum space into regions of short and long wavelengths, 
delimited by an intermediate cutoff $(\omega_0,\boldsymbol{q}_0)$
with $\omega_0\gtrsim \omega$ and $|\boldsymbol{q}_0|\sim \omega_0^\frac{1}{3-x}$.
~\cite{footnote: scale separation} 
The contributions to the self-energy or to the vertex functions
can then be decomposed in short  and long wavelength parts,
following, for example, the procedure of the constructive renormalization.~\cite{Feldman95}
For instance,
\begin{equation}
\Sigma(\omega,\boldsymbol{k}) =
\Sigma^{<}(\omega,\boldsymbol{k})+\Sigma^{>}(\omega,\boldsymbol{k}),
\end{equation} 
where all the internal gauge propagators in the Feynman diagrams 
contributing to the short wavelength part $\Sigma^{>}(\omega,\boldsymbol{k})$
of the self-energy, carry a momentum larger than $(\omega_0,\boldsymbol{q}_0)$. 
By the Ward identities, the self-energy $\Sigma^{>}(\omega,\boldsymbol{k})$,
which is a regular function of $\omega$ and $\boldsymbol{k}$, 
determines the renormalization by the short wavelength contributions of both
the Green function and the vertex function.

The Feynman diagrams that contribute to the remaining part $\Sigma^{<}(\omega,\boldsymbol{k})$
of the self-energy, referred to as the long wavelength part of the self-energy,
contain at least one gauge line carrying a small momenta.
If one of the gauge line carrying a small momenta is singled out,
the long wavelength contribution $\Sigma^{<}(\omega,\boldsymbol{k})$
can be expressed in the form 
\begin{equation}
\begin{split}
\Sigma^{<} & (\omega,\boldsymbol{k}) = 
\frac{i}{m^2} \int^< \frac{d\Omega}{2\pi} \int^< \frac{d^2 k'}{(2\pi)^2} 
D_{\mu\mu}(\Omega,\boldsymbol{k}-\boldsymbol{k}')
\\
&\times
\Gamma_\mu(\omega,\boldsymbol{k};\omega-\Omega,\boldsymbol{k}')
\Gamma_\mu(\omega-\Omega,\boldsymbol{k}';\omega,\boldsymbol{k})
G(\omega-\Omega,\boldsymbol{k}'),
\end{split}
\label{eq: long wl self-energy}
\end{equation}
where $\Gamma_\mu$ is the dressed vertex function.
In Eq.~(\ref{eq: long wl self-energy}), 
the integration range is restricted such that the frequency $\Omega$ and the transfer momenta 
$q=|\boldsymbol{k}-\boldsymbol{k}'|$ are small.

Suppose that the leading contribution to the long wavelengths self-energy 
comes from its transversal part $\Sigma_\perp^{<}(\omega,\boldsymbol{k})$,
involving only the transversal vertex function $\Gamma_\perp$.
(This assumption will be justified below.)
The effect of the short wavelength gauge fluctuations 
on the vertices $\Gamma_\perp$ is included in a multiplicative renormalization
constant $Z_{\perp}^{>}$. 
Moreover, by Ward's identities, 
$Z_\perp^{>} = \big(1+\frac{\partial \Sigma^{>}}{\partial \epsilon_{\boldsymbol{k}}}\big)$.
Furthermore, the integration over the internal momentum $\boldsymbol{k}'$
in Eq.~(\ref{eq: long wl self-energy}) yields a factor of 
$\big(1+\frac{\partial \Sigma^{>}}{\partial \epsilon_{\boldsymbol{k}}}\big)^{-1}$.
The gauge propagator, which is determined by the electron's
charge and electron density does not get renormalized 
by short wavelengths fluctuations.~\cite{Stern95}
The leading term for effective mass then becomes
\begin{equation}
\begin{split}
\frac{m^*}{m} & \simeq
\frac{%
1 -Z_\perp^{>}\partial_\omega \widetilde{\Sigma}^{<}_\perp
}%
{%
1+\partial_{ \epsilon_{\boldsymbol{k}}} \Sigma^{>}
+Z_\perp^{>}\partial_{ \epsilon_{\boldsymbol{k}}}\widetilde{\Sigma}^{<}
}%
%\\
%& 
\simeq
- 
\frac{
\partial_\omega \widetilde{\Sigma}^{<}_\perp(0,k_F) 
}
{
1+\partial_{\epsilon_{\boldsymbol{k}}} \widetilde{\Sigma}^{<}_\perp(0,k_F) 
}+\cdots,
\end{split}
\label{eq: effective mass from long wavelengths}
\end{equation}
where $\tilde{\Sigma}_\perp^{<}(\omega,\boldsymbol{k})$ is the self-energy obtained
from Eq.~(\ref{eq: long wl self-energy}) replacing the vertex functions
$\Gamma_\perp$ by their long wavelength counterpart $\Gamma_\perp^{<}$.
Observe that only the long wavelength contributions to the self-energy
participate in the renormalization of the most divergent part of the effective mass.

%%%%%%%%%%%%%%%%%%%%%%%%%%%%%%%%%
\subsubsection{Long wavelength contributions}

We now argue that including the long wavelength vertex corrections does not modify
the nature of the power-law (respectively the logarithm for $x=1$) 
divergence of the effective mass. 
For this purpose, we show that the self-energy 
$\Sigma^{<}(\omega,\boldsymbol{k})$ behaves in the same
way as the one-loop self-energy~(\ref{eq: one-loop self-energy, all x}) does,
using an induction over the number of loops.
Suppose that the leading term of the long wavelength self-energy 
has been computed to the $n-$loop order and is, up to subleading corrections,
\begin{equation}
\Sigma^{<}_{n}(\omega,k)-\Sigma^{<}_{n}(0,k_F)\simeq
F_n(x)\frac{m_{\mathrm{int}}}{m} 
\left(\frac{|\omega|}{\omega_{\mathrm{int}}}\right)^{2/(3-x)}
\label{eq: n-loops self-energy}
\end{equation}
with a complex function $F_n(x)$. 
The self-energy $\Sigma^{<}_n(\omega,\boldsymbol{k})$ 
can be put into the form of Eq.~(\ref{eq: long wl self-energy}),
with vertex functions $\Gamma_\mu^m$ to a lower order $m<n$ in the loop expansion.

As described in the beginning of this section,
the substitution of the bare propagator $G_0(\omega,\boldsymbol{k})$
by the random-phase-approximation (RPA) 
propagator~(\ref{eq: RPA propagator}) in the sunset diagram
of Eq.~(\ref{eq: long wl self-energy}) can be achieved using a self-consistent
equation for the self-energy, without affecting the leading term of the 
self-energy. 
We should, therefore, replace $G(\omega,\boldsymbol{k})$ 
in Eq.~(\ref{eq: long wl self-energy}) by the bare propagator for this discussion,
considering only the skeleton diagrams for the self-energy. 

The leading contribution to the self-energy 
$\Sigma^{<}_{n+1}(\omega,\boldsymbol{k})$ to the next order 
is then obtained replacing one of the vertex function $\Gamma_\mu^m$
by its expression to the next order in the loop expansion.
We then repeat the calculation that led to the one-loop approximation
for the self-energy. We compute the self-energy at 
$|\boldsymbol{k}|=k_F$, neglecting the $k'$ dependency of the vertex functions
that have no pole for $k'\to k_F$.
Again, integrating over the fermion propagator
restricts the frequency integration to the range $[0,|\omega|]$.
We need an approximation for the vertex functions in the limit
of small frequency and transfer momenta; this can be achieved using
the Ward identities
\begin{subequations}
\begin{eqnarray}
\mathop{\lim_{\Omega \to 0}}_{|\boldsymbol{q}|\lesssim \Omega} 
\Gamma^n_0(\omega,\boldsymbol{k};\omega+\Omega,\boldsymbol{k}+\boldsymbol{q}) & \simeq &
1-\frac{\partial\Sigma^{<}_{n}(\omega,\boldsymbol{k})}{\partial\omega}
\label{eq: vertex function for Omega->0}
\\
\mathop{\lim_{|\boldsymbol{q}|\to 0}}_{\Omega\lesssim |\boldsymbol{q}|} 
\Gamma^n_\perp (\omega,\boldsymbol{k};\omega+\Omega,\boldsymbol{k}+\boldsymbol{q}) & \simeq &
1+ \frac{\partial\Sigma^{<}_{n}(\omega,\boldsymbol{k})}{\epsilon_{\boldsymbol{k}}}.
\label{eq: vertex function for q->0}
\end{eqnarray}
\end{subequations}
where $\Gamma^n_0$ and $\Gamma_\perp^n$ are the vertex functions to the $n-$loop order,
up to subleading contributions coming from the short wavelength part of the
self-energy.
Inserting the ansatz~(\ref{eq: n-loops self-energy}) for the self-energy leads to
the limiting behavior $\Gamma_0^n\sim |\omega|^{\frac{x-1}{3-x}}$
and $\Gamma_\perp^n\sim const$.

Armed with these observations, we derive an approximation for the self-energy 
$\Sigma^{<}_{n+1}(\omega,\boldsymbol{k})$ to the next order in the loop expansion. 
We first consider the longitudinal part of the self-energy.
The longitudinal gauge propagator is $D_{00}\simeq 2\pi/m$ 
in the limit's $\omega\to 0$ and $q\to 0$.
The dominant contribution comes from the region of the momentum space
where the vertex function diverges, i.e., when $v_F q\lesssim |\omega|$. 
Collecting the negative powers of $\omega$ coming from the two
vertex functions~(\ref{eq: vertex function for Omega->0}), and
the factor $\sim \omega^2$ coming from the integrals over $\Omega\in[0,\omega]$ 
and $q\lesssim |\omega|/v_F$ leads to
\begin{equation} 
\Sigma^{<}_{0,n+1}(\omega,k_F)-\Sigma^{<}_{0,n+1}(0,k_F)
\sim |\omega|^{4/(3-x)}+\cdots,
\end{equation} 
which is subleading as compared to the ansatz~(\ref{eq: n-loops self-energy}) 
for the $n-$loop self-energy.

We turn to the transversal part of the self-energy. 
According to the Ward identities~(\ref{eq: vertex function for q->0}),
the vertex function can be approximated by a constant in the
small frequencies and long wavelength limit. 
The integration then works in the same way as for the one-loop calculation,
and we recover the ansatz of Eq.~(\ref{eq: vertex function for q->0}),
with a modified function $F_{n+1}(x)$.%
We, thus, believe that the effective mass of the composite fermions is given by
\begin{equation}
\frac{m^*}{m_{\mathrm{int}}} \simeq
F(x)
\lim_{\omega\to 0} \left(\frac{\omega}{\omega_{\mathrm{int}}}
\right)^{(x-1)/(3-x)}
+\cdots,
\label{eq: effective mass short range 3}
\end{equation}
where $m_{\mathrm{int}}$ was defined in Eq.~(\ref{eq: definition of m_int}).

We briefly discuss the case of Coulomb interaction
and show that the effective mass~(\ref{eq: one-loop result for effective mass x=1})
obtained at one loop is exact.
The original argument of Stern and Halperin~\cite{Stern95}
is based on formula~(\ref{eq: effective mass from long wavelengths}),
and the fact that for Coulomb interaction, 
the leading contribution to the effective mass comes from a small range of frequencies,
but from a large range of momenta $q\in [q_0,2k_F]$, with $q_0\sim \sqrt{|\omega|}$.
It follows that the leading
contribution to $\widetilde{\Sigma}^{<}_\perp(\omega,\boldsymbol{k})$
comes solely from the one-loop diagram:
According to Ref.~\onlinecite{Stern95}, 
the coefficient of the divergent term in 
$\partial_\omega\widetilde{\Sigma}^{<}_\perp(0,k_F)$
is not affected if the diagrams 
having two or more lines carrying long wavelength momenta $\boldsymbol{q}$ 
are omitted. Thus, up to subleading contributions
\begin{equation}
\partial_\omega \widetilde{\Sigma}^{<}_\perp(0,k_F) 
=
\left.\partial_\omega \Sigma(0,k_F)\right|_{\mathrm{one loop}}.
\end{equation}
Since $\partial_{\epsilon_{\boldsymbol{k}}}\widetilde{\Sigma}_\perp^{<}(0,k_F)$
is subleading, the effective mass~(\ref{eq: one-loop result for effective mass x=1})
is exact.

%%%%%%%%%%%%%%%%%%%%%%%%%%%%%%%%%%%%%%%%%%%%%%%%%%%%%%%%%%
\section{Evaluation of the Energy Gap}
\label{sec: Evaluation of the Energy Gap}

In the mean-field picture described in Sec.~\ref{sec: The Fermion-Chern-Simons Theory},
the composite fermions experience a residual magnetic field $\Delta B =B/(2p+1)$
at the filling factor $\nu=\frac{p}{2p+1}$.
The fractional quantum Hall effect can then be viewed as an 
integer quantum Hall effect for the composite fermions filling 
exactly $p$ Landau levels.
The mean-field energy gap is then given by the separation between
the Landau levels in the magnetic field $\Delta B$.
In this approximation, the energy scale is set by the cyclotron
energy of the electrons in the magnetic field $B$.
However, the energy gap is expected to be determined by interaction effects;
in particular, it should remain finite after projection into the lowest
Landau level, i.e., in the limit $m\to 0$.
In order to achieve this requirement, one can replace by hand the
band mass of the electrons by
the ``interaction'' mass $m_{\mathrm{int}}$ defined in Eq.~(\ref{eq: definition of m_int}) 
and obtained equating the composite fermions kinetic 
energy $k_F^2/m_{\mathrm{int}}$ with the interaction energy $\omega_{\mathrm{int}}$
defined in Eq.~(\ref{eq: definition of interaction energy}).
In this modified mean-field approximation, the energy gap is given by
\begin{equation}
\begin{split}
E_{\mathrm{mf}}(p) & =
\frac{\Delta B}{e m_{\mathrm{int}}}
%\\
%& = 
=\frac{2\pi \omega_{int}}{(2p+1)}.
\label{eq: mean-field energy gap}
\end{split}
\end{equation}
In particular, the modified mean-field theory predicts that the energy 
gap decays as $1/(2p+1)$ for $p\to \infty$, for all types of interactions. 

We have seen in the last section that the effect of the gauge fluctuations cannot
be neglected. Halperin {\it et al.}~\cite{HLR} proposed
to account for these fluctuations, replacing the interacting mass 
$m_{\mathrm{int}}$ by the effective mass at half-filling $m^*(\omega)$ 
evaluated at the energy gap $\omega= E_g(p)$.
This leads to the following equation
for the energy gap:
\begin{equation}
E_g(p) =
\frac{\Delta B}{e m^*\big(E_g(p)\big)}.
\label{eq: HLR conjecture B}
\end{equation}
In the next section, we will discuss the solution of this equation 
in the one-loop approximation for different interactions.

%%%%%%%%%%%%%%%%%%%%%%%%%%%%%%%%%%%%%%%%%%%%%%%%%%%%%%%%%%
\subsection{
Energy gap from the HLR-conjecture
}

The results of the Sec.~\ref{ssec: Effective mass}
combined with the HLR-conjecture~(\ref{eq: HLR conjecture B})
allow to evaluate the energy gap of the quantum Hall states 
of the main Jain series with filling factors $\nu=\frac{p}{2p+1}$.

We first consider the case of Coulomb interaction.~\cite{Stern95,Morf02} 
Inserting the one-loop result~(\ref{eq: one-loop result for effective mass x=1})
for the effective mass in the gap equation~(\ref{eq: HLR conjecture}), 
one obtains for $\phi=2$,
\begin{equation}
\Delta\omega_c^* =
\frac{2\pi}{2p+1}
\frac{1}{
-\ln \big(\Delta\omega_c^*\big) +i\frac{\pi}{2}
+C_1
},
\label{eq: gap equation Coulomb}
\end{equation}
where we have introduce the energy gap in units of the interaction energy,
\begin{equation}
\Delta\omega_c^* =
\frac{E_g(p)}{\omega_{\mathrm{int}}}.
\end{equation}
The interaction energy $\omega_{\mathrm{int}}$ is defined
in Eq.~(\ref{eq: definition of interaction energy}); for the Coulomb interaction
and $\phi=2$,
$\Delta\omega_c^*= 4\varepsilon E_g(p)/\big(k_Fe^2\big)$.
The constant $C_1$ cannot be estimated within the composite 
fermion approach and included short wavelength contributions as well. 

Observe that the solution of Eq.~(\ref{eq: gap equation Coulomb}) is complex. 
The unphysical imaginary part of the energy gap is a consequence of
the approximations used to derive the gap equation.
The real part of the solution 
of the gap equation~(\ref{eq: gap equation Coulomb}) can be accurately
approximated by~\cite{Stern95}
\begin{equation}
E_g(p) \simeq
\frac{k_F e^2}{\varepsilon}
\frac{\pi/2}{(2p+1)
\big[\ln(2p+1)+C'\big]
},
\label{eq: energy gap x=1}
\end{equation}
with a constant $C'$ that depends on $C_1$.
The imaginary part of the solution is subleading,
\begin{equation}
\Im \Delta\omega_c^* \sim 
\frac{1}{(2p+1)\ln^2(2p+1)}.
\end{equation}
Moreover, the leading contribution to the effective mass obtained
in the one-loop approximation is exact.~\cite{Stern95}
We may, thus, conclude that the one-loop approximation~(\ref{eq: energy gap x=1}) 
for the energy gap is asymptotically exact for $p\to \infty$.

\begin{figure}
\begin{center}
%FILE coulomb_gap.pdf
%\includegraphics[width=0.45\textwidth]{coulomb_gap.pdf}
%\includegraphics[width=0.45\textwidth]{FIGURE/coulomb_gap.pdf}
%\includegraphics[width=0.45\textwidth]{FIGURE/coulomb_gapLabel.pdf}
\epsfig{file=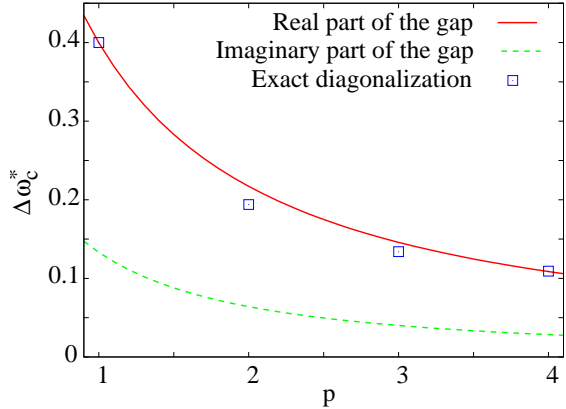,width=0.45\textwidth}
\end{center}
\caption{(Color online)
The energy gap of quantum Hall states at filling factors $\nu=p/(2p+1)$
as function of $p$, for systems with Coulomb interaction.
The energy gap is obtained solving numerically the
gap equation~(\ref{eq: gap equation Coulomb}),
with $C_1=3.8$ (plain line); 
the opposite of the (unphysical) imaginary part of the gap, 
giving an estimation of the quality of the one-loop approximation, 
is also plotted (dashed line).
The dots represent the energy gaps for $p=1,2,3$ and 4
calculated numerically by Morf {\it et al.}~\cite{Morf02}
from the energy gaps obtained by exact diagonalization in finite-size systems.
The energy gap is given in units of $\omega_{int}=k_F e^2 /(4\varepsilon)$.
}
\label{fig: morf02}
\end{figure}

The energy gaps of quantum Hall states at filling factors $\nu = \frac{p}{2p+1}$
with $p=1,2,3$ and $4$ have been estimated from finite-size calculations
by Morf {\it et al.}~\cite{Morf02}
They then used formula~(\ref{eq: energy gap x=1}) in order to fit
their numerical results, using $C'$ as a fitting parameter.
They found that the $p$ dependency of the energy gap is better
explained by Eq.~(\ref{eq: energy gap x=1})
than it would be by the mean-field results of Eq.~(\ref{eq: mean-field energy gap}).
This is surprising, since
the theoretical results are expected to be valid for $p\gg 1$ only.
The estimation for the energy gap from the exact diagonalization of Ref.~\onlinecite{Morf02}
is plotted together with a numerical solution of Eq.~(\ref{eq: gap equation Coulomb})
in Fig.~\ref{fig: morf02}.
The constant $C_1$ is chosen such that the solution of the gap equation
matches the numerical result of Ref.~\onlinecite{Morf02} for $p=1$.

%FIGURE: morf02

We turn to systems with short range interactions ($x<1$). 
Using the one-loop result~(\ref{eq: effective mass short range 2})
for the effective mass,
we obtain from the HLR conjecture~(\ref{eq: HLR conjecture B}) 
the self-consistent gap equation (for $\phi = 2$)
\begin{equation}
\Delta\omega_c^* = 
\frac{2\pi}{2p+1} 
\left(
F(x) \big(\Delta\omega_c^*\big)^{(x-1)/(3-x)}
+\frac{2^{x}}{x-1}
+C
\right)^{-1}
\label{eq: HLR gap equation x<1}
\end{equation}
with an undetermined constant $C$ that depends on $x$. %
(See Fig.~\ref{fig: short, Coulomb, long}.)
For $p\gg 1$, the gap equation~(\ref{eq: HLR gap equation x<1}) 
has the asymptotic solution
\begin{equation}
\Delta\omega_c^* \simeq 
Y(x) \left(  \frac{2\pi}{2p+1} \right)^{(3-x)/2},
\label{eq: HLR gap for x<1}
\end{equation}
where the coefficient $Y(x)$ cannot be determined by one-loop calculations. 
It is argued in Sec.~\ref{ssec: Beyond the one-loop approximation}
that the one-loop approximation gives the right exponent for the 
effective mass as function of $\omega$. We, thus, believe that 
Eq.~(\ref{eq: HLR gap for x<1}) gives the right $p$ dependency of the 
energy gap.
In particular, for $x<1$, the energy gap vanishes faster than
the mean-field result as $p\to \infty$. 

Using the HLR conjecture~(\ref{eq: HLR conjecture B}) and
the one-loop approximation for the effective mass, we
obtain $Y_0(x)=F_0^{\frac{x-3}{2}}(x)$, i.e., the solution~(\ref{eq: HLR gap for x<1})
is complex. Moreover, the ratio between the imaginary part and the real part of the 
solution~(\ref{eq: HLR gap for x<1}) is independent of $p$, for $p\gg 1$.
This indicates that the one-loop approximation only gives a qualitative estimation
of the energy gap as function of $p$.

The asymptotic solution~(\ref{eq: HLR gap for x<1}) of the gap equation 
is only valid as long as
\begin{equation}
(x-1)\ln |\Delta\omega_c^*| \gtrsim 2.
\label{eq: criterium for gap}
\end{equation}
For $x$ close to 1, and $p<\infty$, this inequality might be violated
and the constant term $2/(x-1)$ on the right-hand side of Eq.~(\ref{eq: HLR gap equation x<1})
becomes important. In this case, the gap equation reads
\begin{equation}
\Delta\omega_c^* \simeq
\frac{2\pi}{2p+1}
\left[
\frac{2}{x-1}\left(1- \big(\Delta\omega_c^*\big)^{(x-1)/2}\right)
+i\frac{\pi}{2}+C_1
\right]^{-1},
\label{eq: HLR gap for x->1}
\end{equation}
up to terms of order $(x-1)$.
Observe that the leading term [proportional to $1/(x-1)$]
in this equation is \textit{exact}, i.e., is not modified
by contributions from higher orders in the loop expansion,
as in the case of Coulomb interaction. 
In particular, a solution of Eq.~(\ref{eq: HLR gap for x->1})
should be used instead of the asymptotic solution~(\ref{eq: HLR gap for x<1})
for comparison with exact diagonalization.

For short range interactions ($x>1$), the effective mass $m^*$ is finite and real.
The HLR-conjecture~(\ref{eq: HLR conjecture B}), thus, predicts for $p\gg 1$
the energy gap 
\begin{equation}
\Delta\omega_c^* \simeq \frac{2\pi C'}{2p+1},
\end{equation}%
with an undetermined constant $C'$ (See Fig.~\ref{fig: short, Coulomb, long}). 
In particular,
the energy gap falls off as function of $p$ according to the mean-field prediction.

%FIGURE: morf02
\begin{figure}
\begin{center}
%FILE coulomb_gap.pdf
%\includegraphics[width=0.45\textwidth]{short_Cb_long_real_gap.pdf}
%\includegraphics[width=0.45\textwidth]{FIGURE/short_Cb_long_real_gap.pdf}
\epsfig{file=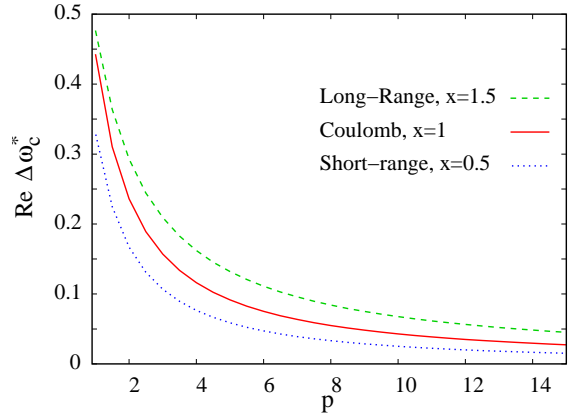,width=0.45\textwidth}
\end{center}
\caption{(Color online)
The energy gap of quantum Hall states at filling factors $\nu=p/(2p+1)$ 
as function of $p$ for 
short range interaction $V(r)\sim 1/r^{3/2}$ (dotted line),
Coulomb interaction $V(r)\sim 1/r$ (plain line), and 
long range interaction $V(r) \sim 1/\sqrt{r}$ (dashed line).
The gaps are obtained solving the HLR-gap equation~(\ref{eq: HLR conjecture B});
only the real part of the solution is plotted. 
The energy gap is given in units of $\omega_{int}=k_F^2 \hat{V}(k_F)/(4\pi\phi)$. %
}
\label{fig: short, Coulomb, long}
\end{figure}

%%%%%%%%%%%%%%%%%%%%%%%%%%%%%%%%%%%%%%%%%%%%%%%%%%%%%%%%%%
\subsection{Direct evaluation of the energy gap}
\label{ssec: Direct evaluation of the energy gap} 

In this section we evaluate the energy gap
of the quantum Hall states at filling factors $\nu=\frac{p}{2p+1}$ directly,
using the Chern-Simons composite fermion approach.~\cite{Stern95}
Gauge fluctuations will be included to the one-loop order.
We will, in particular, verify the conjecture~(\ref{eq: HLR conjecture}) 
of Halperin {\it et al.}~\cite{HLR} for short  and long range interactions.

The starting point is the action defined in Eq.~(\ref{eq: FCS partition function})
for the composite fermions in a weak magnetic field,
in which case the propagator for the composite
fermions is not diagonal in momentum space. 
However, following Stern and Halperin,~\cite{Stern95}
the propagator can be diagonalized in the Landau levels' representation:
As proposed by Haldane,~\cite{Haldane83}
the problem of a two-dimensional electrons gas in a magnetic field 
perpendicular to the  plane can be studied, considering 
a system on a sphere with a magnetic field perpendicular to the surface.
In the spherical geometry, the states in the $n$th Landau level are eigenstates
of the angular momentum operator $\boldsymbol{L}^2$.
States within a Landau levels can then be distinguished as being different eigenstates
of the operator $L_z$.
By conservation of the angular momentum, the propagator for the 
composite fermions on a sphere of infinite radius is diagonal,
and has the general form
\begin{equation}
G(\omega,l) =
\frac{1}{\omega - \epsilon(l)-\Sigma(\omega,l)+i0_+\sgn(\omega) },
\label{eq: LL full fermion propagator}
\end{equation}
where $l$ labels the Landau levels in the residual magnetic field $\Delta B$.
The dispersion relation is defined by
\begin{equation}
\epsilon(l) = 
E_0\cdot  (l+1/2)-\epsilon_F(p),
\quad
E_0 = 
\frac{\Delta B}{e m},
\end{equation}
and $\epsilon_F(p)$ is the chemical potential for which exactly $p$ Landau levels
are filled, and such that adding a fermion to the system would 
start filling the $p+1$th Landau level.
Observe that we work at \textit{fixed} $\Delta B$,
such that the cyclotron frequency $E_0=\omega_c/(2p+1)$ is independent of $p$, 
as discussed in Ref.~\onlinecite{Stern95}.
By definition, the propagator~(\ref{eq: LL full fermion propagator}) 
has a pole at $\omega=0$ for $l=p$
corresponding to the Fermi level. Hence,
\begin{equation}
\epsilon_F(p) =
E_0 \cdot \big(p+1/2\big)+\Sigma(0,p).
\end{equation}
The next Landau level corresponds to a pole in 
Eq.~(\ref{eq: LL full fermion propagator}) located at 
$l=p+1$ and $\omega=E_g(p)$, where
\begin{equation}
\begin{split}
E_g(p) & =
E_0\cdot \big(p+3/2\big) -\epsilon_F(p)+\Sigma\big(E_g(p),p+1\big)
\\
& =
E_0
+\Sigma\big(E_g(p),p+1\big)-\Sigma\big(0,p\big).
\label{eq: full gap equation}
\end{split}
\end{equation}
As discussed in Ref.~\onlinecite{Stern95},
the quantity $E_g(p)$ is the energy gap of the fractional quantum Hall state
at $\nu=\frac{p}{2p+1}$. 
In the limit $\omega_{\mathrm{int}}/\omega_c\to 0$, which corresponds to the projection
of the original electrons gas into the lowest Landau level,
and for $p\gg 1$, the gap equation can be approximated by
\begin{equation}
\begin{split}
\Sigma\big(E_g(p),p\big)-\Sigma\big(0,p\big) & \simeq
-\frac{\omega_c}{2p+1},
\label{eq: gap equation}
\end{split}
\end{equation}
where we have assumed $\partial_{\epsilon(p)}\Sigma(\omega,p)\simeq 0$. 
We now compute the self-energy $\Sigma(\omega,l)$ for $l\simeq p$ and
$\omega$ small, and will show that the gap equation~(\ref{eq: gap equation})
leads to the same equation as derived from the HLR-conjecture~(\ref{eq: HLR conjecture B}),
up to subleading terms.
 
%%%%%%%%%%%%%%%%%%%%%%%%%%%%%%%%%%%%%%%%%%%%%%%%%%%%%%%%%%
\subsection{Evaluation of the self-energy}
\label{ssec: Evaluation of the self-energy}

We now evaluate the self-energy in the residual magnetic field $\Delta B$
at the one-loop order.
As discussed in Ref.~\onlinecite{Stern95}, the RPA-gauge propagator $D_p(\Omega,q)$
in the presence of the residual magnetic field $\Delta B$ 
can be approximated by its limiting form~(\ref{eq: RPA gauge propagator}) 
at half-filling, i.e., for $p\to \infty$. This approximation is justified since
the main contribution to the one-loop self-energy
comes from transfer momenta satisfying $q> k_F/(2p+1)$.
[See Eq.~(\ref{eq: range of momenta}).]
In this formalism, the transversal self-energy is
\begin{equation}
\begin{split}
\Sigma_\perp(\omega,l) = 
i\sum_{j\ge 1}\int & \frac{d\Omega}{2\pi}\int\frac{d^2 q}{(2\pi)^2}
M_{lj}(|\boldsymbol{q}|)\\
& \times G_0(\omega-\Omega,j) D(\Omega,|\boldsymbol{q}|).
\end{split}
\end{equation}
The matrix elements for the transversal current operator are
\begin{equation}
M_{lj}(|\boldsymbol{q}|) =
\frac{1}{N_\phi} \sum_{g,g'}
\left| \big\langle lg \big|J_\perp(\boldsymbol{q})\big| lg' \big\rangle  \right|^2,
\end{equation}
where $l$ and $j$ label Landau levels and 
$N_\phi$ is the degeneracy of the Landau levels labeled by $g$ and $g'$.
The matrix elements $M_{pp}(|\boldsymbol{q}|)$ are evaluated in 
the Appendix~\ref{app sec: The current matrix elements}.

As at half-filling, we compute the difference $\Sigma(\omega,l)-\Sigma(0,l)$
rather than the self-energy directly.
In the limit of large $p$, most contributions to the sum over the Landau levels
come from $j\simeq p$. Since the matrix elements depend smoothly on $j$,
we may neglect their $j$ dependency and set $M_{jl}(q)\simeq M_{pp}(q)$ for $l\simeq p$
and $p\gg 1$.
Further, we approximate the sum over $j$
by an integral
that is performed closing the contour in the upper complex half-plane;
the result of the integration is to restrict the frequency integration to 
$\Omega\in[0,\omega]$. Performing the frequency integration, we obtain
for the transversal part of the self-energy,
\begin{equation}
\begin{split}
\Sigma_\perp(\omega,l)-\Sigma_\perp(0,l)  \simeq 
-
\frac{k_F^2 }{E_0^{\ }} 
\int_0^\infty\!\!\! du \, u^2 M_{pp}\big(k_F u\big)
H(u),
\end{split}
\label{eq: self-energy finite p A}
\end{equation}
where we have defined
\begin{equation}
\begin{split}
H(u) & =
\frac{i}{2}
\ln
\left[
1+\left(u^{x-3}  \frac{\omega}{\omega_{\mathrm{int}}}\right)^2
\right]
+
\arctan
\left[ 
u^{x-3}
\frac{\omega}{\omega_{\mathrm{int}}}
\right],
\end{split}
\end{equation}
with the dimensionless momentum $u=q/k_F$.
As explained in Appendix~\ref{app sec: The current matrix elements},
the matrix elements $M_{pp}(q)$ are non-negligible
only for momenta satisfying
\begin{equation}
q/k_F \in \left[(2p+1)^{-1} ,1\right].
\label{eq: range of momenta}
\end{equation}
In this range, we have from Eq.~(\ref{app eq: M_pp matrix element}) 
\begin{equation}
M_{pp}(q) \simeq
\frac{k_F^2}{\pi m^2 }\frac{k_F}{(2p+1) q}.
\label{eq: matrix element, final}
\end{equation}
Inserting this approximation in Eq.~(\ref{eq: self-energy finite p A})
gives
\begin{equation}
\begin{split}
\Sigma_\perp(\omega,l) &
-\Sigma_\perp(0,l)  \simeq 
-
\frac{k_F^2}{\pi m }
\frac{
I\big( u_{\mathrm{max}} \big)
-
I\big( u_{\mathrm{min}}\big)
}{u^2_{\mathrm{max}}},
\end{split}
\label{eq: self-energy finite p B}
\end{equation}
where $I(v)$ is defined in Eq.~(\ref{eq: integral I(v)})
and computed in Appendix~\ref{app sec: The integral I(v)}.
We have introduced
\begin{equation}
\begin{split}
u_{\mathrm{max}}  =
\big(\omega_{\mathrm{int}}/\omega\big)^{1/(3-x)}
\quad\mbox{ and }\quad
u_{\mathrm{min}}  =
\frac{u_{\mathrm{max}} }{2p+1}.
\end{split}
\label{eq: self-energy finite p C}
\end{equation}
We are interested in the value of the self-energy at
the energy gap, $\omega=E_g(p)$. Anticipating the results from
the HLR conjecture, we expect $u_{\mathrm{max}}\sim \sqrt{2p+1}$
to be large for $p\gg 1$, while $u_{\mathrm{min}}\sim 1/\sqrt{2p+1}$
is expected to be small.
Inserting the approximations~(\ref{app eq: I(v) for small v})
and~(\ref{app eq: I(v) for large v})
of $I(v)$ for small and large arguments in Eq.~(\ref{eq: self-energy finite p B})
gives at the gap energy $\omega=E_g(p)$ 
with $\Delta\omega_c^* = E_g(p) / \omega_{\mathrm{int}}$
\begin{equation}
\begin{split}
 \Sigma_\perp(E_g(p),p) &
- 
\Sigma_\perp(0,p)  \simeq 
\\
& - 
\frac{k_F^2}{2\pi m }
\left\{
F_0(x)\big(\Delta\omega_c^*\big)^{2/(3-x)}
+\frac{2\Delta\omega_c^*}{x-1} 
+\cdots
%\right. 
%\\ & \left.
%+\mathcal{O}\Big(\frac{\ln(2p+1)}{(2p+1)^2}\Big)
\right\}.
%-i\frac{3-x}{(2p+1)^2}
%\left(
%-\frac{1}{3-x} \ln \Delta\omega_c^*
%-\ln(2p+1)
%\right)
%+\mathcal{O}\Big(
%i\big(\Delta\omega_c^*\big)^2,
%(2p+1)^{-2}
%\Big)
\end{split}
\label{eq: self-energy finite p D}
\end{equation}
Inserting the result 
$\Delta\omega_c^*\sim (2p+1)^{-(3-x)/2}$
expected from the HLR-conjecture indicates that 
the first term in curly brackets on the right-hand side 
of Eq.~(\ref{eq: self-energy finite p D})  
is of order $1/(2p+1)$.
The second term is proportional to $(2p+1)^{(x-3)/2}$ and 
leads to the logarithmic behavior for $x\to 1$.

Inserting the obtained expression for the self-energy
in the gap equation~(\ref{eq: gap equation}) gives
after taking the limit of weak interaction $\omega_{int}/\omega_c\to 0$,
\begin{equation}
\begin{split}
\Delta\omega_c^*
& \simeq
\frac{2\pi}{2p+1}
\left[
F_0(x)\big(\Delta\omega_c^*\big)^{(x-1)/(3-x)}
+\frac{2}{x-1} +\cdots
%\right. \\ & \left. 
%+\frac{i (3-x)}{(2p+1)^2\Delta\omega_c^*}
%\left(
%\frac{\ln \Delta\omega_c^*}{3-x} 
%+\ln(2p+1)
%\right) 
\right]^{-1},
\label{eq: one-loop gap equation, x<1}
\end{split}
\end{equation}
where the dots stand for subleading terms in $p$,
which corresponds to the gap equation~(\ref{eq: HLR gap for x<1}). 

For systems with Coulomb interaction, 
we obtain from Eq.~(\ref{eq: self-energy finite p B})
\begin{equation}
\begin{split}
\Sigma_\perp & \big(E_g(p),l\big) 
-\Sigma_\perp(0,l)  \simeq 
\\ &
-
\frac{k_F^2}{2\pi m }
\left\{
\left(
-\ln \big(\Delta\omega_c^*\big)
+
\frac{i\pi}{2}+C_1
\right)\Delta\omega_c^*
%\right.\\
%&\left.
%+i\frac{\ln\big(\Delta\omega_c^* (2p+1)^2\big)}{(2p+1)^2}
+\cdots
\right\},
\end{split}
\end{equation}
where we have neglected terms of order 
$\ln \Delta\omega_c^*/(2p+1)^2$ and $\ln(2p+1)/(2p+1)^2$.
%$(2p+1)^{-2}$, 
%$\big(\Delta\omega_c^*/\omega_{\mathrm{int}}\big)^2$
%and $(\omega_{\mathrm{int}}/\Delta\omega_c^*)/(2p+1)^4$.
The gap equation is then
\begin{equation}
\Delta\omega_c^*
\simeq
\frac{2\pi}{2p+1}
\left[
-\ln \big(\Delta\omega_c^*\big)
+
\frac{i\pi}{2}+C_1+\cdots
%+i\frac{
%\ln\big(\Delta\omega_c^*(2p+1)^2\big)
%}{\Delta\omega_c^*(2p+1)^2}
\right]^{-1},
\label{eq: one-loop gap equation, x=1}
\end{equation}
where the dots represent subleading terms in $p$.
Again, the gap equation~(\ref{eq: gap equation Coulomb}) is recovered
for large $p$. 

%%%%%%%%%%%%%%%%%%%%%%%%%%%%%%%%%%%%%%%%%%%%%%%%%%%%%%%%%%
\section{Conclusion}

In this work, we have computed the long wavelength, low energy contribution to the 
effective mass of the composite fermions 
at half-filling for different electron-electron interactions.
Our results, obtained in the fermion-Chern-Simons approach, are consistent
with the effective mass obtained previously in the literature.~\cite{HLR,Kim95,Nayak94}

For short range interactions, we find an effective mass 
that diverges following a power-law. 
The diverging imaginary part of the effective mass
at the one-loop level indicates that the 
Fermi liquid picture breakdown~\cite{Kim95,Nayak94} for short range interactions.
Nevertheless, we argued that the power-law behavior found in the one-loop approximation
is exact, i.e., is not modified by higher-order terms.
However, unlike in the case of Coulomb interaction, higher orders contribute 
to the most divergent part of the effective mass.
The logarithmic divergence of the effective mass obtained for the Coulomb
interaction~\cite{Stern95,HLR} is recovered as the interaction tends to the Coulomb interaction.
For long range interactions, the effective mass is finite, and presumably real if 
higher-order contributions are properly included.

In the second part of this work, we have considered the energy gap,
which can be either computed from the effective mass of the composite 
fermions using the HLR conjecture,~\cite{HLR}
or directly within the fermion-Chern-Simons picture in a reduced 
magnetic field $\Delta B$.~\cite{Stern95} 
Both approaches lead to the same self-consistent equation for the energy gap.
For short range interaction, the energy gap tends to zero faster
than the mean-field prediction when the filling factor approaches $1/2$.
Moreover, as for the effective mass, the power-law behavior predicted
at the one-loop level turns out to be exact to all orders in perturbation theory,
but the prefactor of the leading term of the energy gap gets contributions
from all orders in perturbation theory.

At the one-loop level, the computed value for the energy gap
has a large imaginary part for short range interactions.
This imaginary part is unphysical since the quantum Hall states
are gaped, i.e., cannot decay.
Thus, including properly all terms in the perturbation expansion
should lead to the exact energy gap,
which behavior is described by the power-law computed at the one-loop level.

As already pointed out in the introduction,
the leading contributions to the self-energy
arise from strong, long wavelength gauge fluctuations
that are not suppressed at large distances if the interaction is short ranged.
The system is then described by a gas of strongly coupled composite fermions; 
in particular, as predicted by renormalization theory,~\cite{Nayak94}
the perturbative Fermi liquid description of the system is not valid.
In this context, the question whether the HLR conjecture 
under the form of Eq.~(\ref{eq: HLR conjecture})
is valid remains open. 
For long range interaction ($x>1$), the HLR conjecture is
probably exact, since the Fermi liquid behavior at half-filling 
is well established by renormalization analysis;
for Coulomb interaction, the (exact) one-loop results of
Stern and Halperin~\cite{Stern95} confirm this conjecture.
The present work established that for short range interactions,
the HLR conjecture gives the right power-law behavior for the energy gap.
However, the HLR conjecture in the form of Eq.~(\ref{eq: HLR conjecture})
cannot be exact if the effective mass of the composite fermions at
half-filling is not real. In this case, one may have to 
modify the HLR conjecture,~\cite{Halperin06} replacing,
for instance, the effective mass by its real part 
in Eq.~(\ref{eq: HLR conjecture}).
Alternatively, an imaginary cutoff in the effective mass might be
used, the phase of the cutoff being fixed by the requirement
that the solution of the gap equation has to be real.

As established in previous works,~\cite{Stern95,Morf02}
the agreement between the energy gap computed 
in the fermion-Chern-Simons approach with Coulomb interaction,
and the same quantity obtained by exact diagonalization
for the filling factors $\nu=1/3,2/5,3/7$ and $4/9$
is surprisingly good.
The same numerical computations have been repeated
for the same filling factors, but for different interactions
of short  and long ranges by Morf.~\cite{Morf06}
The results for the energy gap are not consistent with our 
predictions. In particular, the energy gap as function of the filling factor
decays slower for the short range interaction, than it does for 
the Coulomb interactions, and slower for Coulomb interaction, than
for the long range interaction, as half-filling is approached from below.%
~\cite{Morf06}

The origin of the discrepancy between the numerical results of Morf~\cite{Morf06}
and our theoretical calculations can have different explanations.
First, recall that our theoretical calculations are 
valid in the long wavelength limit, and for filling factors
asymptotically approaching half-filling ($p\to\infty$).
In exact diagonalization studies, however, small systems
with filling factor up to $\nu = 4/9$, i.e., $p\le 4$,
are considered.
One cannot exclude that, for small $p$'s, the self-energy
is dominated by the short wavelength rather than by the
long wavelength contributions.

Further,  numerical studies of small samples 
may probe mainly the contributions coming from
the short range part of the interaction potential,
which is not accessible within our long wavelength approach,
leading to the observed discrepancies.
In order to circumvent this problem, one might
perform exact diagonalization computations 
for different interaction potentials, all having
the same short range part (up to a distance of order of a few magnetic lengths),
but with different (power-law) behaviors for the long range part.
If numerical studies are oversensitive to the short range part of the interaction,
this procedure would introduce a systematic error, independent of the 
interaction type. 

More general questions concerning the quantum Hall states 
in the vicinity of half-filling remain open.
For instance, the sum of the jumps in the chemical
potential occurring at each filling factor $\nu=p/(2p+1)$
diverges.~\cite{HLR}
This suggests that the Jain series could be interrupted
for some value of $p$; in particular, for short range
interactions, phase separation might occur.
In real sample, phase separation might
be triggered even in the case of long range or Coulomb interactions
by density variations induced by disorder.
However, the FCS approach is probably not appropriated to answer such questions,
which could be studied by Monte-Carlo simulations.

\section*{Acknowledgments}
The author wants to thank R.~Morf for 
his collaboration, and for sharing his numerical results,
C.~Mudry for fruitful discussions, and B.~Halperin
for his helpful comments.

%%%%%%%%%%%%%%%%%%%%%%%%%%%%%%%%%%%%%%%%%%%%%%%%%%%%%%%%%%
\appendix
%%%%%%%%%%%%%%%%%%%%%%%%%%%%%%%%%%%%%%%%%%%%%%%%%%%%%%%%%%
\section{The integral \texorpdfstring{$I(v)$}{I(v)}}
\label{app sec: The integral I(v)}

We shortly evaluate the integral
\begin{equation}
I(v) =
\int_0^{v} du \, u
\left\{
\frac{i}{2} \ln\left[1+u^{2x-6} \right]
+\arctan u^{x-3}
\right\}
\label{app eq: integral I(v)}
\end{equation}
defined in Eq.~(\ref{eq: integral I(v)}). 
In particular, for $x=1$,
\begin{equation}
\begin{split}
I(v) = &
\frac{i}{2}\arctan v^2+
\frac{i}{4}v^2\ln\left[1+v^{-4} \right]
\\ &
+
\frac{v^2}{2}\arctan v^{-2} + \frac{1}{4}\ln\big(1+v^4 \big).
\end{split}
\end{equation}
For small arguments $v\ll 1$, the integral $I(v)$ 
is small since
\begin{equation}
\begin{split}
I(v) \simeq &
-i\frac{3-x}{2}v^2
\left(
\ln v -\frac{1}{2}+i\frac{\pi}{2(3-x)}
\right)
\\
&
+\frac{3-x}{5-x} v^{3-x}
+\mathcal{O}\big(v^{5-x},iv^{8-2x} \big).
\end{split}
\label{app eq: I(v) for small v}
\end{equation}
In the limit $x\to 1$, we obtain, for $v\ll 1$,
\begin{equation}
\begin{split}
I(v) \stackrel{x=1}{\simeq} &
-i v^2 
\left(
\ln v -\frac{1}{2}+i\frac{\pi}{4}+\frac{i}{2}
\right)
+\mathcal{O}\big(v^{4} \big).
\end{split}
\end{equation}
We now turn to large arguments, $v\gg 1$.
In order to expand the integrand of~(\ref{app eq: integral I(v)})
in powers of $u^{x-3}$, we have to split the integration
over $u$ in two regions delimited by an arbitrary constant $V\ge 1$.
For the imaginary part of $I(v)$, we may choose $V\to \infty$,
while for the real part of $I(v)$, we set $V=1$.
The first terms in the expansion in powers of $1/v$ are
\begin{equation}
\begin{split}
I(v) \simeq &
\frac{F_0(x)}{2}
+\frac{v^{x-1} }{x-1} 
+i\frac{v^{2x-4}}{4x-8}
+\mathcal{O}\big(v^{3x-7},iv^{4x-10} \big)
\end{split}
\label{app eq: I(v) for large v}
\end{equation}
where
\begin{eqnarray}
\Re F_0(x) & = &
\frac{\pi(3-x)}{4}
\nonumber
\\
&-&
\sum_{n\ge 0}\frac{4(3-x) (-1)^n}{(2n+1)\big(4-(2n+1)^2(3-x)^2\big)}
\nonumber
\\
\Im F_0(x) & =&
(3-x)\int_0^\infty \frac{u\, du}{1+u^{6-2x}}.
\label{app eq: definition of F0}
\end{eqnarray}
In the limit $x\to 1$, this function tends to
\begin{equation}
F_0(x\to 1)
\simeq -\frac{2}{x-1}+i\frac{\pi}{2}
\end{equation}
Hence, dropping all terms of order $(x-1)$ in Eq.~(\ref{app eq: I(v) for large v})
leads to
\begin{equation}
\begin{split}
I(v) \stackrel{x\to 1}{\simeq} &
\frac{v^{x-1}-1 }{x-1} 
+i\frac{\pi}{4}
-i\frac{1}{4v^2}
+\mathcal{O}\big(v^{-4},iv^{-6} \big)
\\
\stackrel{x\to 1}{=}& 
\ln v 
+i\frac{\pi}{4}
-\frac{i}{4v^2}
+\mathcal{O}\big(v^{-4},iv^{-6} \big).
\end{split}
\end{equation} 

\section{The current matrix elements}
\label{app sec: The current matrix elements}

The matrix elements 
\begin{equation}
M_{ll'}(\boldsymbol{q}) =
\frac{1}{N_\phi} \sum_{g,g'}
\left| \big\langle lg \big|J_\perp(\boldsymbol{q})\big| l'g' \big\rangle  \right|^2
\end{equation}
can be evaluated choosing the $x$ axis in the $\boldsymbol{q}$ direction,
such that the transversal current operator becomes
\begin{equation}
J_\perp(\boldsymbol{q})=v_y e^{i|\boldsymbol{q}| x},
\end{equation}
where $x$ and $v_y$ are the position and velocity operators.
In terms of raising and lowering operators, they have the representation%
~\cite{Cohen-Tanoudji}
\begin{eqnarray}
x & = & 
\frac{1}{\sqrt{2m E_0}} \big( a^+ + a + b^+ + b\big)
\nonumber \\
v_y & = & 
\sqrt{\frac{E_0}{2m}}\big( a^+ + a \big)  
\end{eqnarray}
where $a^+$ and $a$ are the inter-Landau level creation and annihilation operators,
while $b^+$ and $b$ are the intralevel creation and annihilation operators. 
Defining $x_a = (a^++a)/\sqrt{2m E_0}$, we may decompose 
$J_\perp(\boldsymbol{q})$ in the product of
the operators $v_y e^{i|\boldsymbol{q}|x_a}$ and $e^{i|\boldsymbol{q}|x_b}$,
where the inter- and intralevel operators are well separated.
After a simple calculation, we find
\begin{equation}
\begin{split}
M_{ll'}(|\boldsymbol{q}|) 
& =
\left|\big\langle l | v_y e^{i|\boldsymbol{q}| x_a} | l'\big\rangle\right|^2
\\
& =
 E_0^2 \left|\frac{\partial }{\partial |\boldsymbol{q}| }
\big\langle l | e^{i|\boldsymbol{q}| x_a} | l'\big\rangle\right|^2.
\end{split}
\label{app eq: matrix elements bis}
\end{equation}
We need to evaluate the matrix element $M_{pp}(q)$ for large $p$.
Using the a semiclassical approximation for the harmonic oscillator,
\begin{equation}
\big\langle p | e^{i q x_a} | p \big\rangle \simeq
\frac{E_0}{2\pi}\int_0^{\frac{2\pi}{E_0}} dt\, e^{i q R_c \sin (E_0 t)}
= J_0(R_c q),
\end{equation}
where $J_0(x)$ is the zeroth Bessel's function, and 
\begin{equation}
R_c=\sqrt{\frac{2p+1}{m E_0}}
= (2p+1)/k_F 
\end{equation}
is the cyclotron radius in the $p$th Landau level.
The semiclassical approximation is valid only for 
\begin{equation}
q\lesssim \sqrt{m E_0 (2p+1) }
= k_F
.
\end{equation}
For bigger $q$, the matrix element $M_{pp}(q)$ falls off exponentially.
In the range of validity of the semiclassical approximation, 
Eq.~(\ref{app eq: matrix elements bis}) becomes
\begin{equation}
\begin{split}
M_{pp}(q) 
& \simeq
 E_0^2R_c^2 
J_1^2(R_c q).
\end{split}
\label{app eq: matrix elements tierce}
\end{equation}
Using the asymptotic forms of the Bessel's function for small and
large arguments, 
\begin{equation}
J_1(x) \simeq 
\left\{
\begin{array}{ll}
x/2, & x\ll 1
\\
\sqrt{\frac{2}{\pi x}}\cos(x-3\pi/4), & x\gg 1,
\end{array}
\right.
\end{equation}
finally, gives 
(using $E_0 R_c = k_F/m$ and replacing the fast oscillating cosine
by its average value)
\begin{equation}
M_{pp}(q) \simeq
\left\{
\begin{array}{l}
 \frac{k_F^2}{4m^2} (2p+1)^2 (q/k_F)^2 ,
\hfill 
q/k_F\ll 1/(2p+1) 
\\
\frac{k_F^2}{\pi m^2} 
\frac{k_F}{(2p+1) q},
\hfill 
1/(2p+1)\ll q/k_F \lesssim 1
\\
0, \hfill q \ge k_F.
\end{array}
\right.
\label{app eq: M_pp matrix element}
\end{equation}
Observe that this approximation corrects~Eq.~(38) of Ref.~\onlinecite{Stern95};
however, the final result for the energy gap found by Stern and Halperin
was not affected.

\section{Haldane pseudopotential}
\label{app: pseudo-potentials}

The Haldane pseudopotentials~\cite{Haldane83} are 
defined as the interaction energy of electron pair
with given relative angular momentum $m$.

The two-body states in the lowest Landau level
are described by the wave functions,~\cite{MacDonald}
\begin{equation}
\psi_m(z,\bar{z}) =
\frac{\bar{z}^m e^{-z\bar{z}/8\ell^2}}{2^{m+1}\ell^{m+1}\sqrt{\pi m! }},
\end{equation}
where $z=x-iy$ are the relative coordinates of the two electrons,
and $m$ is their relative angular momentum.
The Haldane pseudopotentials are the expectation value 
of the potential energy in a state of fixed angular momentum:
\begin{equation}
\begin{split}
V^{(x)}_m &=
\int d^2r \left|\psi_m(\boldsymbol{r})\right|^2 V_x(\boldsymbol{r}) \\
%& =
%\frac{V_x(2\ell) }{ m!}
%\int_0^\infty dr r^{2m-1+x} e^{-r^2}
%\\
& =
\frac{V_x(2\ell) }{ m!}\Gamma\big(m+\frac{x}{2}\big),
\end{split}
\end{equation}
where $\ell$ is the magnetic length.
For Coulomb repulsion between the electrons,
the Haldane pseudopotentials are
\begin{equation}
V_m^{Cb} =
\frac{e^2}{\varepsilon 2\ell}
\sqrt{\pi} \frac{(2m-1)!!}{2^m m!}.
\end{equation}
It can be easily verified that the interactions $V_x(r)$
for $0<x<3/2$ satisfy to the inequalities
\begin{equation}
\begin{split}
V_{m} & > V_{m+2}\\
V_{m-2}-V_{m} & > V_m-V_{m+2}.
\end{split}
\end{equation}
Moreover, the highest slope $V_1^{(x)}-V_3^{(x)}$ 
is significantly bigger than the second highest slope $V_3^{(x)}-V_5^{(x)}$.
These ``hard-core'' properties~\cite{Wojs}
seems to be necessary to the formation of quantum Hall states.
For the interactions considered, i.e., for $0<x<3/2$, the
occurrences of quantum Hall states at the filling factors 
$\nu = 1/3, 2/5, 3/7$, and $4/9$ have been tested numerically.~\cite{Morf06}

%%%%%%%%%%%%%%%%%%%%%%%%%%%%%%%%%%%%%%%%%%%%%%%%%%%%%%%%%%%%%% 
%  THEBIBLIOGRAPHY
%%%%%%%%%%%%%%%%%%%%%%%%%%%%%%%%%%%%%%%%%%%%%%%%%%%%%%%%%%%%%%


\begin{thebibliography}{99}
\bibitem{footnote: pseudo-potentials}
The Haldane pseudopotentials corresponding to the the interactions described by
Eq.~(\ref{eq: potential r-space}) 
with $0 < x < 3/2$ show similarities with the Coulomb pseudopotential,
see Appendix~\ref{app: pseudo-potentials}.
In particular, quantum Hall states are expected along the main Jain
series for all these interactions, which is confirmed by numerical
studies for $\nu = 1/3, 2/5, 3/7$, and $4/9$.~\cite{Morf06}
\bibitem{footnote: fourier trafo V}
The Fourier transform of the interaction is 
$\hat{V}(k)= \alpha(x) \hat{V}(k_F) k_F^x /|\boldsymbol{k}|^x$,
with $\alpha(x)=\int_0^\infty du\, u^{x-1} J_0(u)$, where
$J_0$ is Bessel's function and $u=r|\boldsymbol{k}|$. 
For $0< x < 3/2$, the integral is well-defined. 
For $x=3/2$, the integral $\alpha(x)$ diverges at large distance, 
while for $x=0$, the integral diverges for $u\to 0$.
We will not consider these cases in the present work.
\bibitem{footnote: short range interaction}
The short range interactions considered in this work are
rather unusual, being of the form $V(r)\sim 1/r^\eta$, where $1<\eta<2$.
Interactions falling off as $1/r^2$ or faster, the short wavelength
part of the gauge propagator has to be taken into account.
The energy scale is then fixed by 
$\omega_{int}=\frac{k_F^2\hat{V}(k_F)}{4\pi\phi}+\frac{k_F^2}{2m}\big(\phi/6-1/\phi\big)$.
\bibitem{footnote: scale separation} 
For the Coulomb interaction, one has to separate between small and large
frequencies, but consider the whole range of momenta. 
This is due to the fact that the divergent contribution to the effective mass
comes from small frequencies, but from a wide range of momenta.
\end{thebibliography}
\end{document}